\relax
\documentclass[letterpaper]{article}
\usepackage{aaai17}
\usepackage{times}
\usepackage{helvet}
\usepackage{courier}
\usepackage{booktabs}     
\usepackage{multirow}     
\usepackage[font={bf}, tableposition=top]{caption}     
\usepackage{amsmath}
\usepackage[square,numbers]{natbib}     
\usepackage{graphicx}
\usepackage{url}
\usepackage[font={small}]{subfig}

\begin{document}
%
\title{A Long-Term Analysis of Polarization on Twitter\thanks{This is a pre-print of a short paper accepted at ICWSM'17. Please cite that version instead.}}
\author{Venkata Rama Kiran Garimella\\Aalto University\\kiran.garimella@aalto.fi \And Ingmar Weber\\Qatar Computing Research Institute, HBKU\\iweber@hbku.edu.qa}

\maketitle

\begin{abstract}
Social media has played an important role in shaping political discourse over the last decade. At the same time, it is often perceived to have increased political polarization, thanks to the scale of discussions and their public nature.

In this paper, we try to answer the question of whether political polarization in the US on Twitter has increased over the last eight years. We analyze a large longitudinal Twitter dataset of 679,000 users and look at signs of polarization in their (i) network -  how people follow political and media accounts, (ii) tweeting behavior - whether they retweet content from both sides, and (iii) content - how partisan the hashtags they use are. 
Our analysis shows that online polarization has indeed increased over the past eight years and that, depending on the measure, the relative change is 10\%-20\%. Our study is one of very few with such a long-term perspective, encompassing two US presidential elections and two mid-term elections, providing a rare longitudinal analysis. 
\end{abstract}

\section{Introduction}


Social media has had a tremendous impact on society by redefining ways in which we get exposed to information. A recent Pew survey found that more than 60\% of Americans get their news from social media \cite{pew2016}. 
While there have been a lot of benefits that social media has brought about, including access to a wealth of knowledge, connections and information,
social media is also hypothesized to encourage the creation of echo chambers, where users reinforce their own viewpoints and discredit the view points they do not agree with.\footnote{\url{https://goo.gl/SWk1SR}}
This can potentially lead to a downward spiral of ever increasing political polarization\footnote{The term \emph{polarization} always refers to \emph{political polarization} in this paper.}, which, in turn, makes it harder to have a fact-based debate and to reach a consensus on controversial issues.


Though a lot of studies have shown the existence of polarization on social media~\cite{conover2011political,adamic2005political}, little analysis has been done on long term trends. Performing a study across several decades, as has been done to demonstrate the increasing polarization in the US House of Representatives~\cite{andris2015rise}, is of course impossible as social media is still a fairly recent phenomenon. However, since Twitter was founded in 2006 and its usage now spans several US presidential elections, we still have potential data for a decade.


The main question we want to answer in this paper is if political polarization has increased over time on social media. 
%
To address this question, we collect data from Twitter related to both social network structure and tweet content for a large set of users (679,000) engaging with US politics.

We define polarization as a tendency to be restricted in terms of obtaining or engaging with political information to one side of the left-vs.-right political spectrum. To avoid drawing conclusions based on a single perspective, we address three questions each using a different type of information. Namely, (i) have users become less likely to follow both sides of the political spectrum, (ii) have users become less likely to retweet both sides, and (iii) have users become less likely to use hashtags shared by both sides.



Our analysis reveals that, according to all three measures, polarization has increased by 10\% and 20\% between 2009 and 2016. 


To the best of our knowledge, this is the first study analyzing political polarization on Twitter over a period of eight years. As it is always easy to get caught up in the heat of the moment, we believe that our study adds a valuable long-term perspective to the evolution of online polarization in the US.

\section{Related Work}
Potentially the first study to describe political polarization in a data-driven manner was work by Adamic et al.\cite{adamic2005political} who point out a clustered structure of hyperlinks between ideologically opposing blogs. Similar analysis on Twitter~\cite{conover2011political} revealed that political polarization exists on Twitter and manifests itself in a way that users endorse (retweet) their own side, but not the opposing side. 
On a similar note, \cite{garimella2016quantifying} show that most polarized discussions on social media have a well-defined structure, when looking at the retweet network. 


From a content perspective, \cite{mejova2014controversy} consider discussion of controversial and non-controversial news over a span of seven months and identify a 
correlation between controversial issues and the use of biased and emotional language. They measure bias using manually curated sets of keywords and emotional language using lexicon dictionaries, like SentiWordNet.


Weber et al.~\cite{weber2013secular} study temporal changes in political polarization in Egypt and present evidence that increases in their hashtag-based measure of polarization precede events of violence in the real world.
In a similar spirit, Morales et al.~\cite{morales2015measuring} study polarization over a period of two months during the death of Hugo Chavez and identify an increase in polarization in conjunction with external events. 
Yardi et al.~\cite{yardi2010dynamic} study the evolution of a gun violence incident on Twitter for two months and show how homophily plays a role in polarizing discussions.
Du et al.~\cite{du2016echo} compare two separate snapshots of a random sample of the Twitter follower network, one taken in June 2016 and one in August 2016. They then observe that, in line with theories on ``triadic closure'',
`new edges are (at least 3-4 times) more likely to be created inside existing communities than between communities and existing edges are more likely to be removed if they are between communities'. Such mechanisms could lead to the increase in polarization that we observe in our study. 

Perhaps the closest to this work is the study by Andris et al.~\cite{andris2015rise} who study the partisanship of the US congress over a long period of time. They find that partisanship in the US congress has been increasing for the past few decades.


Concerning longitudinal studies of social media, Liu et al.~\cite{DBLP:conf/icwsm/LiuKM14} analyze seven years of Twitter data to quantify how the users, their behavior, and the site as a whole have evolved. Their work, however, does not describe aspects particular to political polarization.

\section{Dataset}
Our dataset is collected around a set of public seed Twitter accounts: politicians and media outlets, with known political leaning. From these seed users we then crawl outwards by collecting data for users who follow or retweet the seed users. Details as follows.


\subsection{Seed Accounts}
Our point of departure is a list with two types of polarized seed accounts. The first type consists of presidential/vice presidential candidates and their parties (see the political accounts in Table~\ref{tab:seed_politicians}) for the last eight years. The second type consists of popular media accounts listed in Table~\ref{tab:seed_politicians}. The list of media outlets was obtained from a report by the Pew Research Center on polarization and media habits.\footnote{\url{http://www.journalism.org/2014/10/21/political-polarisation-media-habits/}}

\subsection{Following Users}
For each seed user, we obtained all their followers. The combined set of all followers for all seed accounts gave us a total of 140M users. We estimated the time when a user followed a particular seed account using the method proposed by Meeder et al.\cite{meeder2011we}. This method is based on the fact that the Twitter API returns followers in the reverse chronological order in which they followed and we can lower bound the follow time using the account creation date of a user. So, as at least some of @BarackObama's followers started to follow him right after creating their Twitter account, this leads to temporal bounds for the other followers as well. 
These estimates are reported to be fairly accurate when estimating follow times for users with millions of followers. 
For our analysis, we used all cases with estimated follow dates from January 2009 onwards.

\subsection{Retweeting Users}


For the set of seed politicians, we obtained all their public, historic tweets\footnote{Since the Twitter API restricts us to the last 3200 tweets, we used a public tool to get all historic tweets \url{https://github.com/Jefferson-Henrique/GetOldTweets-python}}. The earliest tweets in this collection date back to 2006. For each collected tweet, we used the Twitter API to collect up to 100 retweets. This gave us a set of 1.3M unique users who retweeted a political entity since 2006. We randomly sampled 50\% of these users (679,000), and used the Twitter API to get 3,200 of their most recent tweets in December 2016. This gave us around 2 billion tweets. 
Though we have tweets dating back to 2007, we only consider tweets from September 2009 onwards in the analysis since the volume for earlier tweets is low. 
%
%
We perform all our subsequent analysis for retweets and hashtag polarization computation on this data.



\begin{table}[ht]
\centering
\caption{\label{tab:seed_politicians}US seed accounts with known political leaning. Top: political candidates and parties. Bottom: partisan media outlets.}
\begin{tabular}{l|c}
\hline
\hspace{1.5cm}	\textbf{Political accounts}   & \textbf{Side}  \\ \hline
barackobama,joebiden,timkaine,hillaryclinton,\\thedemocrats             & left \\  \hline
realdonaldtrump,mike\_pence,mittromney,gop,\\ speakerryan,senjohnmccain,sarahpalinusa & right \\ \hline
\hspace{1.8cm}\textbf{Media outlets}  & \textbf{Side}  \\ \hline
npr,pbs,abc,cbsnews,nbcnews,cnn,usatoday,\\ nytimes,washingtonpost,msnbc,guardian,\\ newyorker,politico,motherjones,slate,\\ huffingtonpost,thinkprogress,dailykos,edshow & left  \\  \hline
theblaze,foxnews,breitbartnews,drudge\_report,\\ seanhannity,glennbeck,rushlimbaugh & right 
\end{tabular}
\end{table}


\section{Experiments}

In our quest to understand long-term polarization trends, we look at three aspects:
(i) If user are now more likely to follow users across the political spectrum, (ii) if they are now more likely to retweet such users, and (iii) if users are now more likely to use hashtags which are used by both sides. 
Here we describe the three types of experiments we performed, related to following and retweeting behavior in the first part, followed by experiments related to hashtags usage.

\subsection{Following and Retweeting Behavior}
To observe changes in the polarization of the following (retweeting) behavior, we wanted to track changes in the probability to follow (retweet) accounts from both sides. 
As, due to sparsity, following (retweeting) only a single user from one of the two sides is not necessarily a strong signal for polarization, we decided to apply a Bayesian methodology. Before observing any evidence, we gave each following (retweeting) user a uniform prior probability to follow (retweet) seed users from either side. Concretely, we used a beta distribution with a uniform prior ($\alpha$ = $\beta$ = 1), where $\alpha$ measures the left leaning and $\beta$ the right leaning. 

Then every follow (retweet) to either side increases the count for that side by +1, basically simulating a repeated coin toss where we are studying the bias of the coin. As the beta distribution is the conjugate prior of the binomial distribution, we might obtain something like $\alpha=4$, $\beta = 2$  for a (mostly) left leaning user. The mean of the beta distribution, and hence the ``leaning'' $l$ of the following (retweeting) user, is defined as $l=\alpha/(\alpha + \beta)$, taking the leftness as the direction of the index. We defined the polarization $p$ as $p = 2 \cdot |0.5 - l|$, giving a measure between 0.0 and 1.0 measuring the deviation from a balanced leaning.

\begin{figure*}[ht]
\begin{minipage}{.33\linewidth}
\centering
\subfloat[]{\label{fig:follow_politician}\includegraphics[width=\textwidth, height=\textwidth, clip=true, trim=0 0 0 0]{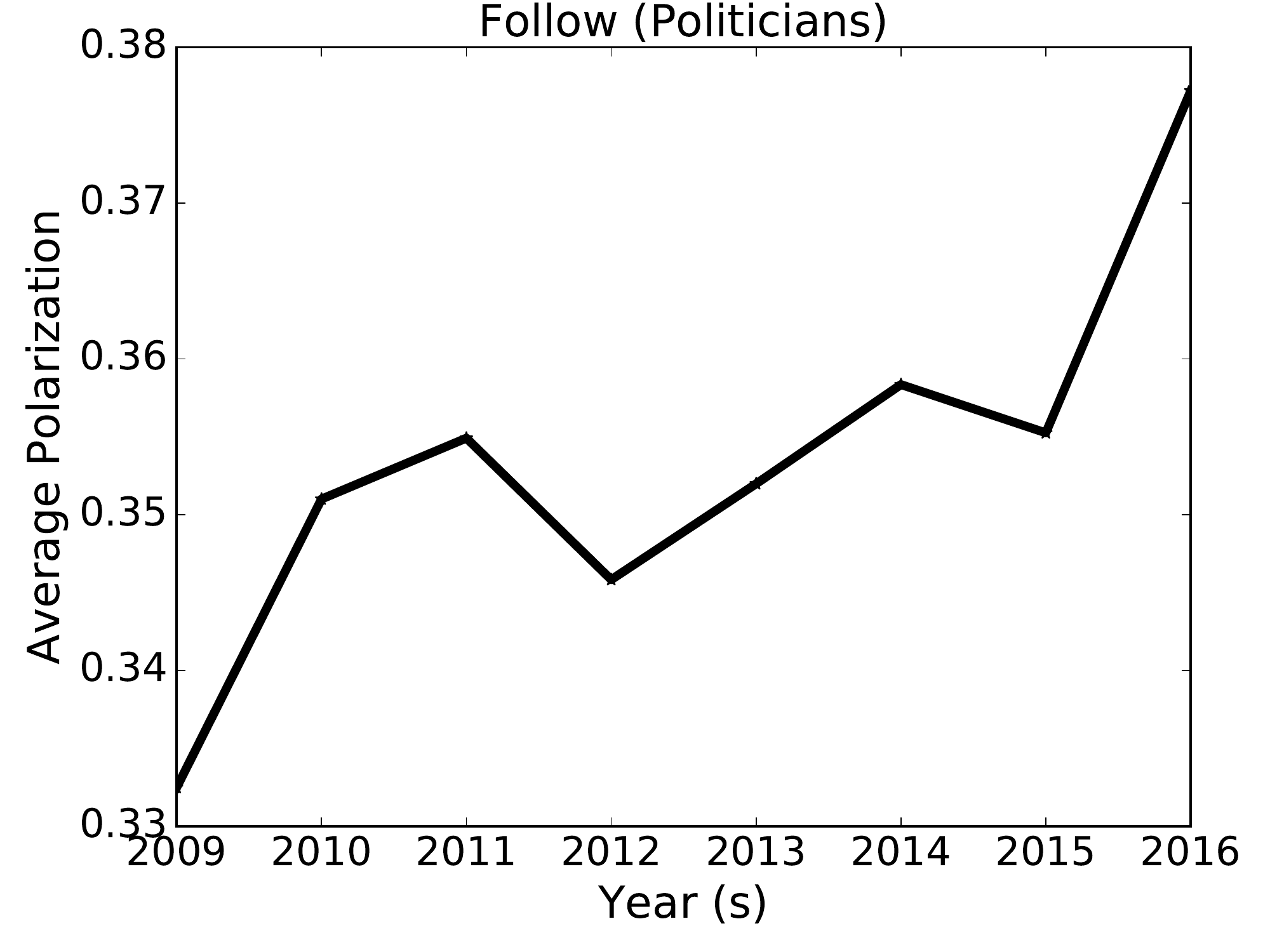}}
\end{minipage}
\begin{minipage}{.33\linewidth}
\centering
\subfloat[]{\label{fig:follow_media}\includegraphics[width=\textwidth, height=\textwidth, clip=true, trim=0 0 5 0]{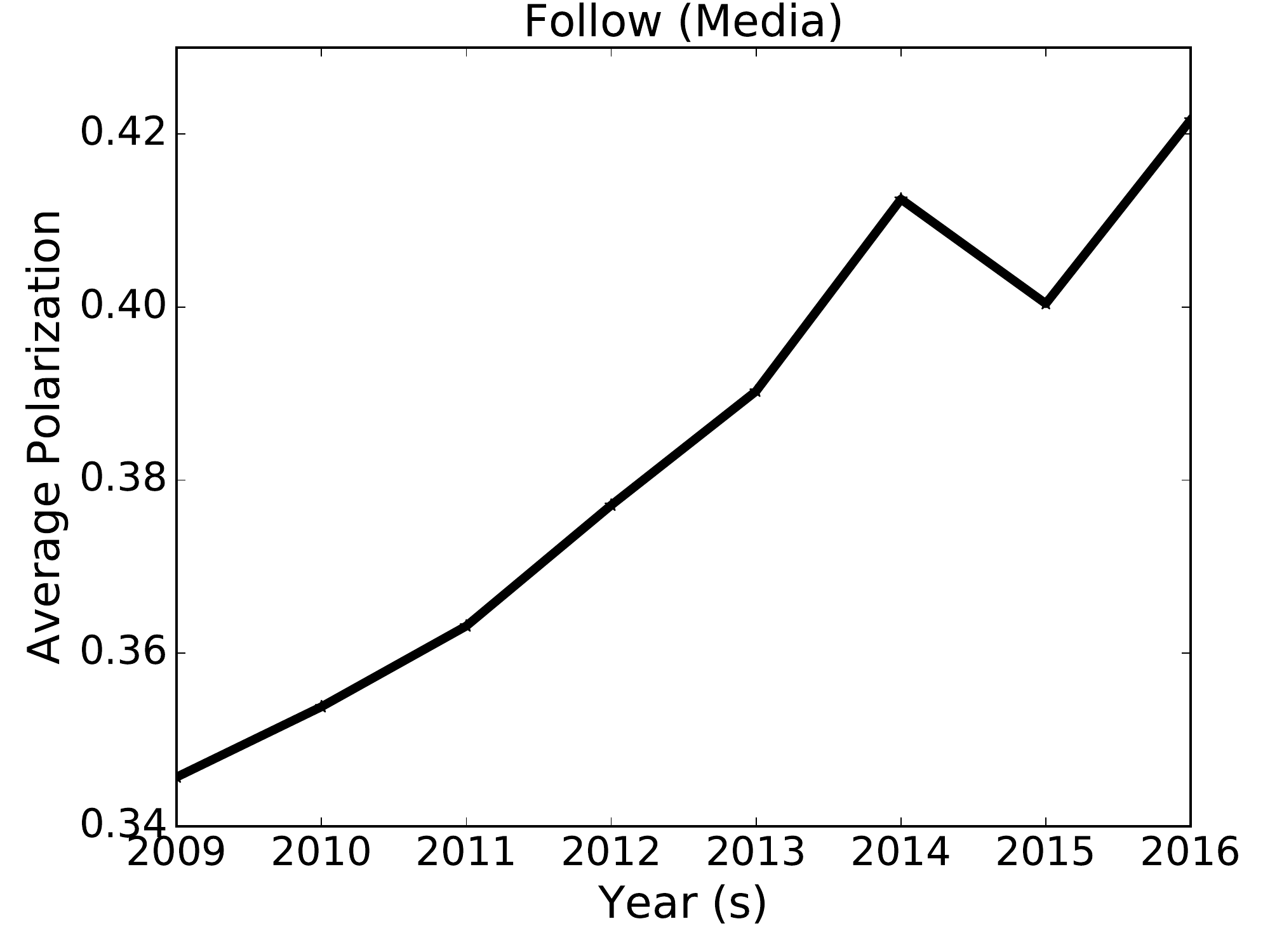}}
\end{minipage}
\begin{minipage}{.33\linewidth}
\centering
\subfloat[]{\label{fig:retweet_politician}\includegraphics[width=\textwidth, height=\textwidth, clip=true, trim=0 0 0 0]{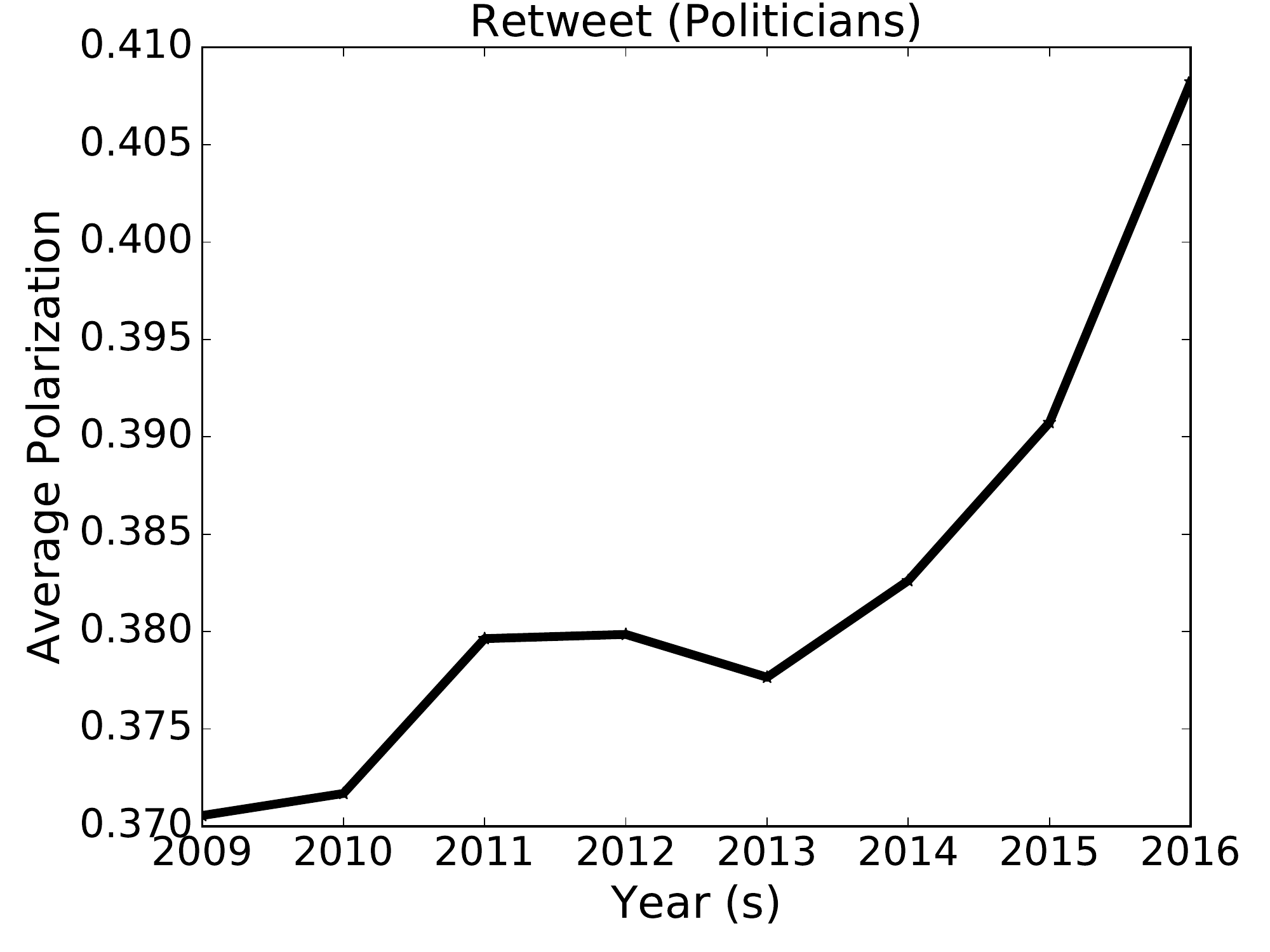}}
\end{minipage}
\caption{Follow (a,b) and retweet (c) effects over time for politicians (a,c) and media (b) seed accounts.}
\label{fig:follow_retweet}
\vspace{-\baselineskip}
\end{figure*}

For each political follower/retweeter and each year, this method gives us a value of polarization. Figure~\ref{fig:follow_retweet} plots the distribution of the average polarization and shows temporal shifts. Note that whereas we have the following information for both political and media seed accounts, we only have the retweeting users for political seed. Regardless of whether using politician or media outlet seed accounts, and regardless of whether using following or retweeting information, polarization has increased from 2009 to 2016 between 10-20\% in relative terms. 




\subsection{Hashtag Polarization}
The third type of polarization we analyze relates to content polarization, more specifically to the polarization of hashtags used by users. Conceptually, a society could be thought of as polarized if there are two opposing sides who speak different languages, in that they differ completely in the words they choose to describe things. For example, one person's ``global warming'' could be another person's ``climategate''. We operationalize this idea by applying the methodology previously used in ~\cite{weber2013political}.

In their methodology, for a given week, a user is assigned a leaning based on the political seed accounts they retweet during that week. Users not retweeting seed accounts during the week do not contribute. Each hashtag $h$ is then assigned a leaning $l_h$ between 0.0=right and 1.0=left based on the leaning of the users using the hashtags in the given week. Differences in user volumes for the two sides are corrected for and smoothing is applied to deal with sparsity. For each hashtag $h$ in a given week, its polarization $p_h$ is then, as before, defined as $p_h = 2 \cdot |0.5 - l|$. The values of $p_h$ are then averaged across all $h$ used in a given week by retweeting users. See \cite{weber2013political} for details.

To further reduce noise due to low volumes, in particular during the early years, we (i) ignored hashtags used by fewer than five users, and (ii) computed moving averages across five weeks.  
To look for drifts in the time series, we first tested for stationarity of the time series. An augmented Dickey-Fuller test found the time series to be non-stationary (p $<$ 0.0001). 
Next, we computed the linear fit across time and tested the value of the non-zero slope for statistical significance using a t-test (p $<$ 0.0001).

\begin{figure*}[ht]
\centering
\includegraphics[width=\textwidth, height=0.4\textwidth, clip=true, trim=10 5 5 5]{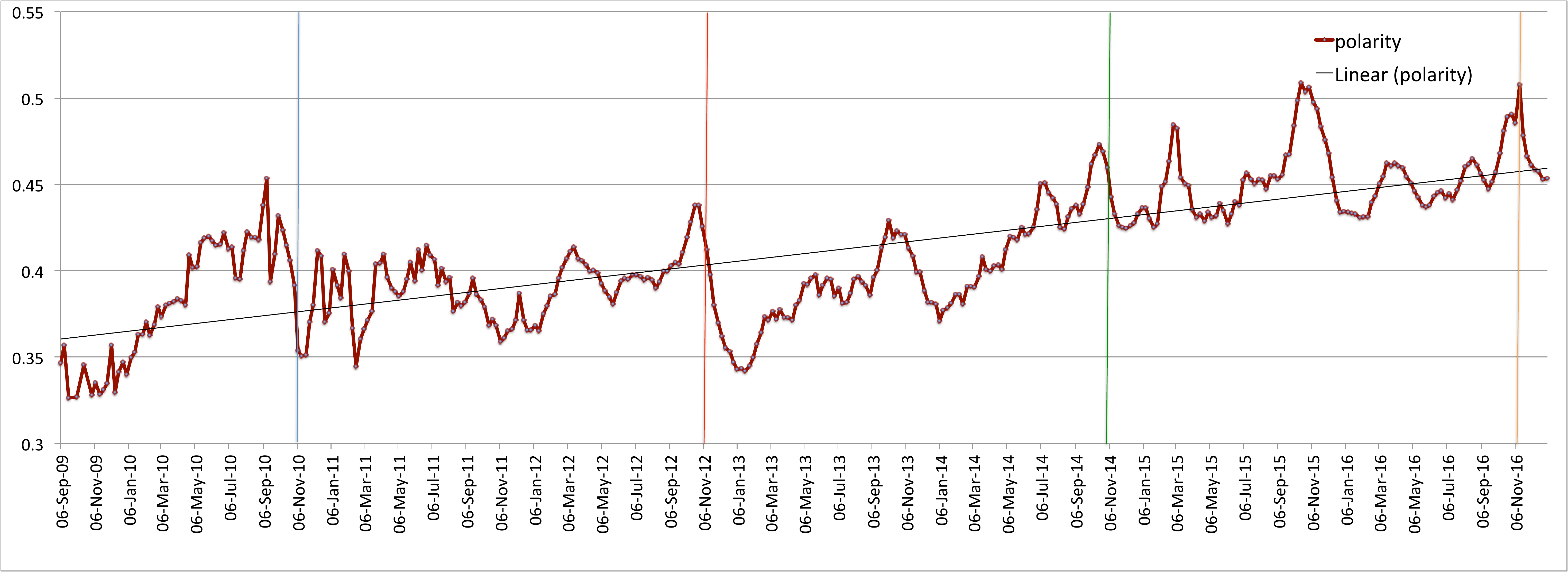}
\caption{\label{fig:polarity} Weekly hashtag polarization. The four vertical lines indicate the 2010 midterm, the 2012 presidential, the 2014 midterm, and the 2016 presidential elections.}
\vspace{-\baselineskip}
\end{figure*}

Figure~\ref{fig:polarity} shows the temporal changes for the measure of hashtag polarization together with the linear fit superimposed. Similar to the following and retweeting polarization, there is a relative increase of about 20\% between 2009 and 2016. 
%
%
Due to the finer time scale, Figure~\ref{fig:polarity} also suggests that the time around elections corresponds to local maxima in polarization, whereas the time after elections corresponds to local minima. For the 2010 midterm elections, however, this observation does not hold, potentially due to noisier estimates based on fewer active users on Twitter.


\section{Conclusions}
There is conflicting evidence on whether social media 
(i) actively increases offline polarization through the formation of online echo chambers, 
(ii) merely reflects offline polarization \cite{vaccari2016echo}, or 
(iii) helps to \emph{reduce} offline polarization by exposing users to a more diverse set of opinions than they would find in their offline social network \cite{bakshy2015exposure}. 

Though our analysis does not directly settle this debate, it provides evidence that polarization on Twitter has increased over the past eight years, potentially reflecting increases in offline polarization as those observed in the US House of Representatives \cite{andris2015rise}. Furthermore, for three different methodologies, the relative size of the increase of polarization was found to be between 10\% and 20\%.

In our work, we did not explicitly attempt to detect ``astroturfing'' and other types of automated tweets \cite{ratkiewicz2011detecting}. However, due to the longitudinal nature of our study, by the time of the data collection (2016/2017) Twitter will have had time to catch most cases of users violating their terms of service, suspending their accounts. More organic efforts such as hashtag hijacking \cite{hadgu2013political} could still affect our analysis, though these effects are arguably also part of the political landscape and should be included. 

Given the running up to the 2016 US presidential elections and concerns of a President Trump - famous for his Twitter politics - doing little to attempt to reduce polarization, we speculate that the online polarization will continue to increase in the foreseeable future. 

Finally, as social media is coming of age and soon teenagers will no longer remember a pre-social-media era, new opportunities for longitudinal studies arise. At the same time, technical challenges related to ``most recent activity only'' API limitations hinder such studies. Still, we expect and look forward to more long-term analysis such as ours in the future.


\vspace{-1mm}
\bibliographystyle{aaai}
\bibliography{biblio}

\end{document}